\documentstyle[pra,aps,amsfonts]{revtex}

\def\beq{\begin{equation}}
\def\eeq{\end{equation}}
\def\beqa{\begin{eqnarray}}
\def\eeqa{\end{eqnarray}}

\begin{document}
\draft

\title{Generalized Faddeev Equations for $N$-Particle Scattering}

\author{G. W. Wei}
\address{Department of Computational Science, National
University of Singapore, Singapore 117543}

\author{R. F. Snider}
\address{Department of Chemistry, University of British Columbia,
Vancouver, British Columbia, Canada V6T 1Z1}
 
\date{\today}
\maketitle

\begin{abstract}

        A proposal is made for reducing the solution of the
$N$-particle Lippmann-Schwinger equation to that of smaller sets of
particles.  This consists of first writing the $N$-particle equation
in terms of all possible $N/2$-particle Lippmann-Schwinger equations.
(If $N$ is odd this needs a minor modification.)  The second step
requires a decoupling of the resolvents for the fewer particle systems
so that each can be solved separately.  This generalization of the
Faddeev approach deals only with connected kernels and the
homogeneous solution reproduces the $N$-particle Schr\"odinger
equation.
For four particles the proposed method involves only a
$3\times3$ matrix whereas other approaches typically require the
solution of at least a $7\times7$ matrix equation. 

\end{abstract}

\pacs{11.80.Jy, 03.80.+r, 11.80.Gw, 21.45.+v}

\section{INTRODUCTION}

        There has been an increased interest in many-body scattering
dynamics due to the rapid growth of computer power\cite{Muriel}. The
fundamental method of formal scattering theory is the
Lippmann-Schwinger equation\cite{LipSch}.  This equation is equivalent
to the Schr\"{o}dinger equation for a two-body problem with an
appropriate scattering boundary condition.  However, it is well
recognized that there are problems when solving the Lippmann-Schwinger
equation for more than two particles.  In particular for three
particles, but the same problem occurs when there are more particles,
a perturbation expansion in powers of the potential, or better in
terms of pair particle T-matrices, allows a third particle to be a
spectator and a dangerous delta function to arise in the momentum
representation of the Lippmann-Schwinger equation \cite{Weinb}.
Faddeev \cite{Faddeev} rearranged the equation so as to eliminate the
dangerous delta function and in the process he was able to show that
the kernel of the rewritten integral equation was compact, essentially
by showing that it became connected after one iteration.
Lovelace\cite{Lovelace} and later, Alt, Grassberger and Sandhas
(AGS)\cite{AGS} gave two different ways of rewriting the Faddeev
equations in terms of scattering amplitudes for various collision
processes. The purpose of this work is to propose a way of 
generalizing the Faddeev equation to an arbitrary number of particles.

          There has been a large effort in extending Faddeev's work to an   
$N$-particle system, in particular to the four-particle system, both  
for theoretical interest and for practical use in nuclear and chemical
reactions
\cite{Weinb,AGS,MGSP,Weyers,Rosenberg,Omnes,Hahn,Glockle1,Yakubovsky,NarYak,Sloan,Bencze,Redish,KarZei,BaeKou,Tobocman,Sandhas,Sandhas2,HabSan}.
Most work puts the emphasis on deriving equations with a connectable  
kernel, namely, a kernel that is connected after a finite number of
iterations.  Gl\"{o}ckle\cite{Glockle1} showed that a set of
generalized Lippmann-Schwinger equations has a unique solution for the
three-particle system.  He also formulated similar equations for a
four-particle scattering system\cite{Glockle2}.  Special formulations
were given by Weinberg\cite{Weinb} and Rosenberg\cite{Rosenberg}, the
former emphasized the quasiparticle picture of an interacting
$N$-particle system, the latter provided a set of equations that do
not explicitly depend on the scattering potentials, a feature that may
be useful for relativistic scattering\cite{AleOmn}.  A channel array
approach was taken by Baer {\it et al.}\cite{BaeKou}, to achieve a
connected kernel for a three-body system which was later generalized
to an $N$-particle scattering system by Tobocman\cite{Tobocman}.
Yakubovsky\cite{Yakubovsky} presented a formal generalization of the
Faddeev equations to $N$-particle scattering by appropriately
organizing pair-particle interactions.  Yakubovsky's equations
\cite{Yakubovsky}, like those of Faddeev, have considerable advantage
in dealing with the bound state problem since they are equivalent to
the Schr\"{o}dinger equation, thus there being less chance of spurious
solutions\cite{Glockle3,Glockle4}.   Unfortunately, the complexity and
the lack of a simple method of iterating the Yakubovsky equations
hinder their practical usage in physical and chemical problems. In a
later treatment, Narodetsky and Yakubovsky\cite{NarYak} proposed an
alternative two-cluster approach  and obtained a new set of equations
for the $N$-particle scattering. The equivalence of these two-cluster
equations to the Schr\"{o}dinger  equation has not be proved. Sloan
made an important contribution to solving the four-particle problem by
allowing only two-body channels in the coupled equations\cite{Sloan}
for scattering amplitudes. Bencze\cite{Bencze} obtained generalized
Lovelace equations and generalized AGS equations by resuming the
two-cluster equations obtained by Narodetsky and Yakubovsky and showed
that these equations are equivalent to those of Sloan when $N=4$.
Redish\cite{Redish} generalized Sloan's equations to an arbitrary
number of particles and demonstrated that his results are identical to
those of Bencze.  The Sloan-Bencze-Redish equations are connected
after one iteration provided that the kernels of the subpartitions are
connected.

        These previously presented methods appear to emphasize the
pair particle transition operator and expand everything in terms of
it. This approach of building up everything from pair particle
properties involves combining the pair transition operators in more
and more elaborate ways, the  details of which may often be very
tedious. In fact, there are so many terms that they are very
cumbersome to implement.  Even the comparison of the numerical
calculations using different methods appears to raise questions.  What
is proposed here might be thought of as an approach of pulling the
problem apart into smaller pieces.  Thus the $N$-particle system
(if  $N$ is even) is divided into all possible
combinations of $N/2$-particle subsystems with the coupling of the
subsystems carried out in a manner parallel to the method of Faddeev.
This is repeated for each $N/2$-particle subsystem.  Eventually, after
successive divisions, each subsystem consists of two particles, whose
collisional effects are described by a pair transition operator. (A
minor modification is required if $N$ is odd.) This gives a structure
as to how the $N$-particle transition operator is dependent on the
transition operator of fewer and fewer particles.  The calculation of
the $N$-particle transition operator then involves the reverse
procedure of successively putting together the transition operators of
fewer particles to eventually get an expression for the full
transition operator.  In going up the chain of partitionings, each
step leads to an integral equation of Faddeev type which is connected
after one iteration.  In this way a set of equations is obtained in
which all particles are connected, but the connectedness is organized
in sets within sets.  A necessary technicality for carrying out this
approach is the need to renormalize the potential at each
partitioning.  Thus the transition operators for smaller sets of
particles involve a potential which is generally smaller in magnitude
than the true potential.  It is believed that the approach presented
here has a different basis of approach than those previously
presented, has  fewer terms and hopefully
leads to a more efficient numerical procedure.

          This paper is divided into five sections.  Section II is
devoted to the partitioning of $N$-particle systems, with emphasis on
partitions having two clusters with an equal number of particles.  The
notation is similar to that of Redish \cite{Redish} and Yakubovsky
\cite{Yakubovsky}.  The distinction between a channel and a partition
is stressed.  A scheme is developed in Section III to generalize the
Faddeev equations to a set of equations for $N$-particles.  The
derivation given for the generalized Faddeev equations closely
parallels Faddeev's original derivation.  This emphasizes the
simplicity of the new equations.  The generalized Faddeev equations
for the resolvent operator and the decomposition of the scattering
wave functions are given in Section IV.  Section V discusses
how the proposed procedure would be applied to both 4- and 8-particle
scattering.   A  general discussion in Section VI ends the
paper.

\section{Properties of Equal Partitioning}
\label{KINEM}

        Consider a system of $N$ labeled particles $(1,2,3,\cdots,N)$.
There are many topologically distinct ways of partitioning the labeled
$N$ particles into sets of clusters of particles.  Each particular way
of dividing up the $N$ particles will be called a partition, labeled
by $A, B, C$, etc.  The detailed presentation given here partitions
the $N$ particles into two  clusters of nearly
equal size (if $N$ is odd, one cluster is to have one more particle
than the other), but alternate partitioning schemes are also
possible, as  mentioned in the Discussion.  All partitionings
of the same size are to be considered, which differ by particle
composition in each cluster, to give a set of partitions $\cal C$.
The objective of this section is to define the potentials and
hamiltonians appropriate for such partitions. 
As used in this work, a partition only becomes a (rearrangement)
channel when the clusters composing the partition are bound, thus
distinguishing between the mathematical method of solving the
$N$-particle problem and the physical notion of what asymptotic states
arise in a scattering process.

        The $N$-particle hamiltonian ${\bf H}={\bf H}_0 +{\bf V}$
consists of a kinetic energy operator ${\bf H}_0$ and an assumed
pairwise additive potential operator
\begin{equation}
{\bf V} \equiv \sum_{i<j} {\bf V}_{ij},
\end{equation}
where the sum is over all possible ordered pairs, with
$i,j\in\{1,2,3,\cdots,N\}$.  The total hamiltonian is assumed to be
self-adjoint and bounded below, and to have spectra that might
include a discrete set, associated with the bound eigenstates, as well
as the continuum for the scattering states.  The objective is to
determine the solutions for the Schr\"odinger equation associated with
the total hamiltonian ${\bf H}$,
\begin{equation}\label{Schro}
{\bf H}\mid \Psi\rangle=E\mid\Psi\rangle.
\end{equation}

       In the present treatment, the $N$ particles are partitioned
into two clusters of equal size if at all possible.  That is, if
$N$ is even, there are $N/2$ particles in each, whereas if $N$ is
odd, one cluster in a partition has $(N+1)/2$ particles and the other
$(N-1)/2$.  It follows that the number of partitions in this set
${\cal C}$ of partitions is given by
\begin{equation}\label{noofpart}
{\cal N}_N=\cases{
{1\over2}{N\choose N/2}\hspace{1cm}&for $N$ even;\cr
N\choose (N-1)/2&for $N$ odd.\cr}
\end{equation}
The partition potentials ${\bf V}_C$ are to be chosen so that
\begin{equation}\label{VN}
\sum_{C\in{\cal C}}{\bf V}_C \equiv {\bf V}.
\end{equation}
Thus the potential ${\bf V}_C$ of partition $C$ is defined as
\begin{equation}
{\bf V}_{C}=v_N\sum_{j<k\atop j,k\in C}{\bf V}_{jk},
\end{equation}
involving the sum over those pairs of particles $jk$ that appear in
the same cluster in the partition $C$. Here $v_N$ is a renormalization
factor and obviously, $v_3=1$. For $N>3$, $v_N$ is chosen as
\begin{equation}\label{scale}
v_N=\cases{{N\choose2}/\left[{N/2\choose2}{N\choose N/2}\right]=
{(N/2)!((N/2)-2)!\over(N-2)!}\hspace{0.5cm}&$N$ even;\cr
{N\choose2}/\left\{{N\choose(N-1)/2}\left[{(N-1)/2\choose2}+
{(N+1)/2\choose2}\right]\right\}=\\
{((N+1)/2)!((N-3)/2)!\over(N-1)!}&$N$ odd,\cr}
\end{equation} 
so that the total number of potential terms on both sides of Eq.
(\ref{VN}) are equal.  The special case of $v_4=1$ implies that no
renormalization is needed for $N=4$, whereas $v_5=1/4$ and $v_6=1/4$
demonstrate how the potential for a particular pair of particles gets
distributed among the many partitions in the set ${\cal C}$ when $N$
is large.  Note that other types of two cluster partitions can also be
selected (or included), which however, leads to a larger number of
coupled equations for a given $N$-particle system.

        A partition hamiltonian is defined as the kinetic energy
operator ${\bf H}_0$ plus the mutual potential interactions between
those particles in each of its clusters, namely the partition
potential, thus
\begin{equation}
{\bf H}_{C}\equiv{\bf H}_0+{\bf V}_{C}.
\end{equation}
The residual interaction, ${\bf V}^{C}$, describing how the clusters
in partition $C$ interact with each other, is defined as
\begin{equation}
{\bf V}^{C}\equiv {\bf V}-{\bf V}_{C}.
\end{equation}
Thus, for each partition $C$ the total hamiltonian ${\bf H}$ can be
written as
\begin{equation}
{\bf H}={\bf H}_{C} + {\bf V}^{C}.
\end{equation}
The resolvent operator ${\bf R}(z)$ for the total hamiltonian is
defined as
\begin{equation}
{\bf R}={1\over z-{\bf H}},
\end{equation}
where $z$ is a complex parameter, for scattering theory having a
small positive imaginary part.  In most equations in this work the
dependence on $z$ will be implicitly assumed.  The free resolvent
operator is defined as
\begin{equation}
{\bf R}_0={1\over z-{\bf H}_0}
\end{equation}
while the resolvent operator for partition $C$ is
\begin{equation}\label{R_C}
{\bf R}_{C} \equiv{1\over z-{\bf H}_{C}}.
\end{equation}
The object of this paper is to express the total transition
operator ${\bf T}$, as determined by the Lippmann-Schwinger equation
\begin{equation}\label{LSeq}
{\bf T}\equiv{\bf V}+{\bf V}{\bf R}_0{\bf T},
\end{equation}
and the Schr\"{o}dinger equation (\ref{Schro}) in terms of the
clustering discussed above.

\section{GENERALIZED FADDEEV EQUATIONS FOR ${\bf T}$}

        In this section a two-step scheme is proposed for the
decomposition of the total transition operator ${\bf T}$ in such a
manner that its kernel is connected after one iteration.  The first
step is to decompose ${\bf T}$ according to the selected set of {\it
two-cluster} partitions ${\cal C}$, see Sec. \ref{KINEM}.  Conceptually,
this is exactly what Faddeev\cite{Faddeev} did for the three-body
system.  However for a system of more than three particles, the
resulting expression for a partition transition operator ${\bf T}_C$
is governed by a combination of the transition operators for the two
clusters.  The second step is to identify how this combination is to
be carried out, compare Ref. \onlinecite{Sloan}.  Finally it is
recognized that the separate cluster transition operators are exactly
similar to the original $N$-particle transition operator but with
significantly fewer particles, so the whole procedure can be repeated
for each cluster.  Care must be exercised when reconstructing the
partition transition operators from the separate cluster transition
operators so that the energies appearing in the resolvents are
consistent with the total energy of the system.

\subsection{The $N$-Particle Transition Operator}

        The full $N$-particle transition operator has the equivalent
forms
\begin{eqnarray}
{\bf T}&=&{\bf V}+{\bf V}{\bf R}{\bf V}\nonumber\\
&=&{\bf V}+{\bf T}{\bf R}_0{\bf V} \nonumber\\
&=&{\bf V}+{\bf V}{\bf R}_0{\bf T}.
\end{eqnarray}
Moreover, it can be decomposed according to
\begin{eqnarray}\label{T_2}
{\bf T} &=&\sum_{C\in {\cal C}} {\bf V}_{C}
+\sum_{C\in {\cal C}} {\bf V}_{C}{\bf R}_0{\bf T} \nonumber\\
&=&\sum_{C\in {\cal C}}{\bf T}^{C},
\end{eqnarray}
where
\begin{equation}\label{FaddT2}
{\bf T}^{C} \equiv {\bf V}_{C}+{\bf V}_{C}{\bf R}_0{\bf T}.
\end{equation}
Subtracting ${\bf V}_{C}{\bf R}_0{\bf T}^{C}$ from both sides of Eq.
($\ref{FaddT2}$) gives
\begin{equation}
\left({\bf 1}-{\bf V}_{C}{\bf R}_0\right){\bf T}^{C} = {\bf V}_{C}
+{\bf V}_{C}{\bf R}_0\left({\bf T}-{\bf T}^{C}\right),
\end{equation}
which, on multiplying by $({\bf1}-{\bf V}_{C}{\bf R}_0)^{-1}$ gives
\begin{eqnarray}\label{fadd2}
{\bf T}^{C}&=&\left({\bf 1}-{\bf V}_{C}{\bf R}_0\right)^{-1}
{\bf V}_{C}\nonumber\\
&&+\left({\bf 1}-{\bf V}_{C}{\bf R}_0\right)^{-1}{\bf V}_{C}
{\bf R}_0\left({\bf T}-{\bf T}^{C}\right)\nonumber\\
&=&{\bf T}_{C}+\sum_{B\in {\cal C}}\bar{\delta}_{C,B}{\bf T}_{C}
{\bf R}_0{\bf T}^{B}.
\end{eqnarray}
Here $\bar{\delta}_{C,B}\equiv1-\delta_{C,B}$ and the 
(2-cluster) partition transition operator ${\bf T}_{C}$ is
\begin{eqnarray}\label{partT}
{\bf T}_{C}&\equiv&
\left({\bf1}-{\bf V}_{C}{\bf R}_{0}\right)^{-1}{\bf V}_{C}\nonumber\\
&=&{\bf V}_{C}+{\bf V}_{C}{\bf R}_{0}{\bf T}_{C}
={\bf V}_{C}+{\bf T}_{C}{\bf R}_{0}{\bf V}_{C} \nonumber\\
&=&{\bf V}_{C}+{\bf V}_{C}{\bf R}_{C}{\bf V}_{C}.
\end{eqnarray}
Therefore the whole $N$-particle transition operator ${\bf T}$ can be
split into ${\cal N}_N$ Faddeev type transition operators ${\bf T}^C$
associated with the selected set ${\cal C}$ of two-cluster partitions.
It is interesting to note that Eq. (\ref{fadd2}) has the precise 
structure of the Faddeev equation but is valid for an arbitrary number
of particles and with ${\cal N}_N$ components.  Just like the original
Faddeev equation, the kernel in Eq. (\ref{fadd2}) is connected after
one iteration provided the scattering kernels of the subsystems are
already connected.

        As in the other treatments\cite{Sloan,Bencze,Redish} of the
$N$-particle system, 2-cluster transition operators ${\bf T}_C$ may be
expressed in terms of unconnected subsystems.  Thus it is necessary to
ensure that each partition transition operator is connected in order
for the total transition operator to have a connected kernel.  This
can be accomplished by assuring that each cluster transition operator
is connected.  But first, it is necessary to know how to express the
partition transition operators in terms of the corresponding 
single cluster transition operators.  This is done in the next
subsection.

\subsection{Cluster transition operators}

        The partition transition operator ${\bf T}_C$ can be
calculated from the transition operators for the two clusters $C_1$
and $C_2$ that constitute the partition $C$. Two approaches for
carrying out calculations of this nature have been presented in the
literature.  First is the approach of Sloan \cite{Sloan} for a
4-particle system, see his Sec. III. This is essentially similar to
the method of the last subsection, dividing up the partition
transition operator into two parts as in Eq. (\ref{T_2}) and obtaining
coupled equations for the two parts, as in Eq. (\ref{fadd2}).  The
second approach is to separate the partition resolvent into cluster
resolvents by the use of a convolution.  This method is useful since
the two clusters are dynamically independent, so all operators for one
cluster commute with all operators of the other cluster.  This
property of the resolvent was pointed out by Bianchi and Favella
\cite{BF} and emphasized for use in parts of the 4-particle problem by
Haberzettl and Sandhas \cite{HabSan}.  Quantities that are used by
both methods are defined first, then the methods are discussed in
turn.

        On the basis that the two clusters in a partition are
dynamically independent, the partition potential ${\bf V}_C$ is a sum
of cluster components
\beq
{\bf V}_C={\bf V}_{C_1}+{\bf V}_{C_2}.
\eeq
Since the kinetic part, ${\bf H}_0$, of the hamiltonian also separates
into cluster components, it follows that the partition hamiltonian
also separates into commuting cluster hamiltonians.  That is, these
hamiltonians are related according to $(j=1,2)$
\beq
{\bf H}_0={\bf K}_{C_1}+{\bf K}_{C_2},~~~~
{\bf H}_{C_j}={\bf K}_{C_j}+{\bf V}_{C_j},~~~~
{\bf H}_C={\bf H}_{C_1}+{\bf H}_{C_2}.
\eeq
In a similar manner, cluster transition operators can also be defined
according to 
\beqa\label{Tcluster}
{\bf T}_{C_j}(z')&\equiv&{\bf V}_{C_j}+{\bf V}_{C_j}
{1\over z'-{\bf K}_{C_j}}{\bf T}_{C_j}(z')\nonumber\\
&=&{\bf V}_{C_j}+
{\bf T}_{C_j}(z'){1\over z'-{\bf K}_{C_j}}{\bf V}_{C_j}\nonumber\\
&=&{\bf V}_{C_j}+{\bf V}_{C_j}{1\over z'-{\bf H}_{C_j}}{\bf V}_{C_j}.
\end{eqnarray}
What complex parameter $z'$ is to appear in each cluster transition
operator depends on how it is to be used, and differs between the two
methods.

        In the first method, modelled on Sloan's \cite {Sloan}
approach, the partition transition operator is written as a sum,
\beq\label{Tchannel}
{\bf T}_C={\bf T}^{C_1}+{\bf T}^{C_2},
\eeq
whose parts are defined as
\begin{equation}\label{FaddT2-2}
{\bf T}^{C_j} \equiv {\bf V}_{C_j}+{\bf V}_{C_j}{\bf R}_0{\bf T_C}.
\end{equation}
Then repeating a process analogous to deriving Eq. (\ref{fadd2}), the
components of the partition transition operator can be shown to
satisfy the coupled equations
\begin{equation}\label{Tchannel2}
\left(\begin{array}{c} {\bf T}^{C_1}\\
{\bf T}^{C_2}\end{array} \right)=
\left(\begin{array}{c} {\bf T}_{C_1}(z_1)\\
{\bf T}_{C_2}(z_2)\end{array}\right)+
\left(\begin{array}{cc} 0 & {\bf T}_{C_1}(z_1) \\
{\bf T}_{C_2}(z_2) &0 \end{array} \right){\bf R}_0
\left(\begin{array}{c} {\bf T}^{C_1}\\
{\bf T}^{C_2}\end{array} \right),
\end{equation}
where the parameters
\beq
z_1=z-{\bf K}_2,\hspace{2cm}z_2=z-{\bf K}_1
\eeq
have been chosen so that the energy factors are consistent with the
properties of the partition transition operator.  According to this,
the partition transition operator can be expressed as the series
\beq\label{Tchannel3}
{\bf T}_C={\bf T}_{C_1}(z_1)+{\bf T}_{C_2}(z_2)+{\bf T}_{C_1}(z_1)
{\bf R}_0{\bf T}_{C_2}(z_2)+{\bf T}_{C_2}(z_2){\bf R}_0
{\bf T}_{C_1}(z_1) +\cdots.
\eeq
The cluster transition operators ${\bf T}_{C_j}(z_j)$ can be evaluated
for  arbitrary $z_j$ entirely in the respective cluster subspace
involving the states of ${N\over 2}$ or ${N-1\over 2}$ particles in
$C_j$.  This can be accomplished in the same manner as that described
for the $N$ particle system since the cluster is dynamically
independent.  But it must be evaluated with the appropriate $z_j$
parameter when used in calculating the partition transition operator.

        In the second method the basic starting point is to express
the partition resolvent as the convolution
\beq
{\bf R}_C={-1\over2\pi i}\int_\Gamma{dz'\over(z'-{\bf H}_{C_1})
(z-z'-{\bf H}_{C_2})}
\eeq
of the resolvents of the two clusters.  Here the contour $\Gamma$ is
to be the straight line from $-\infty$ to $\infty$, but lying above
the real axis and below $z$.  Thus $z'$ must be such that
$0<\Im(z')<\Im(z)$.  It follows that the partition transition operator
is given by
\beq\label{TCT1T2}
{\bf T}_C={\bf V}_{C_1}+{\bf V}_{C_2}+{-1\over2\pi i}\int_\Gamma
({\bf V}_{C_1}+{\bf V}_{C_2}){dz'\over(z'-{\bf H}_{C_1})
(z-z'-{\bf H}_{C_2})}({\bf V}_{C_1}+{\bf V}_{C_2}).
\eeq
This can be expressed in many different ways.  The following
emphasizes how the integrand can be expressed in terms of the cluster
transition operators, Eq. (\ref{Tcluster}), but leaves the contour
integral unchanged.

        The integrand consists of four terms, according to the
different cluster potentials.  The diagonal in cluster potential
terms can immediately be recognized as related to the corresponding
cluster transition operator, whereas the remainder needs the identity
\beq
{\bf V}_{C_1}{1\over z'-{\bf H}_{C_1}}={\bf T}_{C_1}(z')
{1\over z'-{\bf K}_{C_1}}
\eeq
and its other various combinations.  After some calculation, the
integrand can be written in the form
\beqa\label{integrand}
&&\hspace{-1cm}({\bf V}_{C_1}+{\bf V}_{C_2}){1\over(z'-{\bf H}_{C_1})
(z-z'-{\bf H}_{C_2})}({\bf V}_{C_1}+{\bf V}_{C_2}) \nonumber \\
&=&[{\bf T}_{C_1}(z')-{\bf V}_{C_1}]{1\over z-z'-{\bf H}_{C_2}}+
[{\bf T}_{C_2}(z-z')-{\bf V}_{C_2}]{1\over z'-{\bf H}_{C_1}}\nonumber\\
&&+{\bf V}_{C_1}{1\over(z'-{\bf H}_{C_1})(z-z'-{\bf H}_{C_2})}
{\bf V}_{C_2}+{1\over z'-{\bf H}_{C_1}}{\bf V}_{C_1}{\bf V}_{C_2}
{1\over z-z'-{\bf H}_{C_2}} \nonumber \\
&=&[{\bf T}_{C_1}(z')-{\bf V}_{C_1}]{1\over z-z'-{\bf K}_{C_2}}+
[{\bf T}_{C_2}(z-z')-{\bf V}_{C_2}]{1\over z'-{\bf K}_{C_1}}\nonumber\\
&&+[{\bf T}_{C_1}(z')-{\bf V}_{C_1}]{1\over z-z'-{\bf K}_{C_2}}
{\bf T}_{C_2}(z-z'){1\over z-z'-{\bf K}_{C_2}} \nonumber \\
&&+[{\bf T}_{C_2}(z-z')-{\bf V}_{C_2}]{1\over z'-{\bf K}_{C_1}}
{\bf T}_{C_1}(z'){1\over z'-{\bf K}_{C_1}} \nonumber \\
&&+{\bf T}_{C_1}(z'){1\over(z'-{\bf K}_{C_1})(z-z'-{\bf K}_{C_2})}
{\bf T}_{C_2}(z-z') \nonumber \\
&&+{1\over z'-{\bf K}_{C_1}}{\bf T}_{C_1}(z'){\bf T}_{C_2}(z-z')
{1\over z-z'-{\bf K}_{C_2}}.
\eeqa
Most terms involve the product of the transition operators for the two
clusters, weighted with different combinations of the free particle
resolvents for the clusters.  It is up to the computational method to
decide which way these are to be evaluated.  But the first two terms
each involves  only one cluster transition operator. The contour
integral of these two terms can easily be done.  For the first term
the combination ${\bf T}_{C_1}(z')-{\bf V}_{C_1}$ is analytic in the
upper half $z'$-plane and vanishes for $|z'|\to\infty$, so closing
the contour at $\Im(z')\to+\infty$ contributes a pole only when
$z'=z-{\bf K}_{C_2}$.  Analogously for the second term, so that the
partition transition operator can be written
\beq\label{Tconv}
{\bf T}_C={\bf T}_{C_1}(z_1)+{\bf T}_{C_2}(z_2)
+{-1\over2\pi i}\int_\Gamma~~\left\{{\bf T}_{C_1}
{\bf T}_{C_2}~~{\rm terms}\right\}.
\eeq
While this is only one way of organizing this result there are many
other ways in which the cluster transition operator expansion of the
partition transition operator could be written and no attempt is made
here to catalog all the possibilities. In comparing the two methods,
Eq. (\ref{Tchannel3}) is an infinite series in cluster transition
operators, while Eq. (\ref{Tconv}) is only quadratic in cluster
transition operators, but requires an integration over how the energy
is divided up between the two clusters.

        In this way the calculation of the partition transition
operator has been reduced to the independent computation of the
cluster transition operators.  The advantage of this decomposition is
that the individual cluster transition operators deal with isolated
sets of particles whose number is less than $N$.  Such a cluster of
particles can be decomposed into partitions as was the original
problem, and the whole process repeated.  Specifically, for $N=2^n$
even, the successive problems deal with $2^k$ particles,
$k=n-1,~n-2,~...,1$.

\section{GENERALIZED FADDEEV COMPONENTS}

        As a function of $z$, the resolvent operator ${\bf R}$ has
singularities at the spectrum of the system hamiltonian, thus it is of
practical use for finding solutions of the Schr\"odinger
equation\cite{Newton}, for determining the dynamical evolution of the
system\cite{CMM,Haberzettl,MWS}, and identifying normalizable
resonance states\cite{WeiSni}.  Faddeev's approach\cite{Faddeev} is
followed in order to express the $N$-particle resolvent operator in
terms of selected two-cluster partition transition operators.  This
representation of the total resolvent operator is then used to
decompose a scattering wave function originating from a particular
incoming two-cluster partition.  The resulting wave functions are here
referred to as generalized Faddeev components.  The homogeneous
system of equations associated with the generalized Faddeev
components is shown explicitly to solve the Schr\"{o}dinger equation.
This section first describes the generalized Faddeev equations for the
$N$-particle resolvent operator and subsequently discusses the
associated wave functions.

        The total resolvent operator ${\bf R}$ is related to the total
transition operator ${\bf T}$ via the Lippmann-Schwinger equation
\begin{eqnarray}\label{R_n1}
{\bf R}&=&{\bf R}_0 + {\bf R}_0{\bf V}{\bf R}
={\bf R}_0+{\bf R}_0{\bf T}{\bf R}_0.
\end{eqnarray}
On taking over the partition expansion (\ref{T_2}) of ${\bf T}$, this
can be written as
\begin{equation}\label{FaddR1}
{\bf R}={\bf R}_0+\sum_{C\in{\cal C}}{\bf R}^{C},
\end{equation}
where the Faddeev type resolvent ${\bf R}^C$ is given by
\begin{equation}
{\bf R}^{C}\equiv {\bf R}_0{\bf T}^{C}{\bf R}_0.
\end{equation}
The generalized Faddeev equations (\ref{fadd2}) lead to the coupled
set of equations
\begin{eqnarray}\label{FaddR2}
{\bf R}^C &=&{\bf R}_0{\bf T}_{C}{\bf R}_0
+\sum_{B\in {\cal C}}\bar{\delta}_{C,B}{\bf R}_0{\bf T}_{C}
{\bf R}_0{\bf T}^{B}{\bf R}_0 \nonumber\\
&=&{\bf R}_{C} - {\bf R}_0
+\sum_{B\in {\cal C}}\bar{\delta}_{C,B}{\bf R}_0{\bf T}_{C}
{\bf R}^{B}
\end{eqnarray}
with partition resolvent operator defined in (\ref{R_C}) and related to
the partition transition operator by
\begin{equation}
{\bf R}_C={\bf R}_0+{\bf R}_0{\bf V}_C{\bf R}_C
={\bf R}_0+{\bf R}_0{\bf T}_C{\bf R}_0.
\end{equation}
For $N=3$, Eq. (\ref{FaddR2}) is Faddeev's decomposition of the total
resolvent operator whose kernel is not connected  until after one
iteration.  For $N>3$, the kernel of Eq. (\ref{FaddR2}) requires one
iteration in the same manner providing the kernels of the subsystems
are already connected.  These relations between the transition
operators and the resolvent operators are analogous to the original
Faddeev equations for the three-particle resolvent operator.  The set
of resolvent equations can also be used for the evaluation of
statistical mechanical virial coefficients\cite{WeiSni,WeiSni2}.

        A scattering wave function for the $N$-particle system is
determined by the total resolvent operator ${\bf R}$ according to
\begin{equation}\label{Wavef1}
\mid\Psi_{C} \rangle =\lim_{\varepsilon\rightarrow 0} i\varepsilon
{\bf R}(E+i\varepsilon)\mid \phi_{C}\rangle.
\end{equation}
This is applied here to the selected set ${\cal C}$ of compatible
two-cluster partitions, with $\mid \phi_{C}\rangle$ a stationary
solution of the two-cluster partition hamiltonian ${\bf H}_{C}$ of
energy $E$, 
\begin{equation}\label{qq}
{\bf H}_{C}\mid \phi_{C}\rangle = E \mid \phi_C\rangle,
\end{equation}
which may be a distorted wave of the two-cluster subsystem.

        The detailed structure of the scattering wave function
$\mid\Psi_{C}\rangle$ of Eq. (\ref{Wavef1}) is now discussed.  This is
begun by first applying the resolvent expansions (\ref{FaddR1}) and
(\ref{FaddR2}) to Eq. (\ref{Wavef1}),
\begin{eqnarray}\label{Wavef2}
\mid\Psi_{C} \rangle &=&\lim_{\varepsilon\rightarrow 0} i\varepsilon
{\bf R}_0(E+i\varepsilon)\mid \phi_{C}\rangle
+\sum_{B\in {\cal C}}\lim_{\varepsilon\rightarrow 0}
i\varepsilon{\bf R}^{B}(E+i\varepsilon)\mid \phi_{C}\rangle.
\end{eqnarray}
The first term on the right hand side contributes only if 
$C$ is the channel with all particles free, namely
\begin{equation}
\lim_{\varepsilon\rightarrow 0} i\varepsilon
{\bf R}_0(E+i\varepsilon)\mid \phi_{C}\rangle = \mid
\phi_{C}\rangle\delta_{C, C_N},
\end{equation}
where $C_N $ is the $N$-cluster partition.  The second term can be
written as a sum of generalized Faddeev components
\begin{equation}\label{Wavef3}
\lim_{\varepsilon\rightarrow 0}i\varepsilon{\bf R}^{B}
(E+i\varepsilon)\mid \phi_{C}\rangle\equiv\mid \psi_{BC}\rangle.
\end{equation}
According to Eq. (\ref{FaddR2}), the generalized Faddeev component 
$\mid \psi_{BC}\rangle$ can be expanded as
\begin{eqnarray}\label{Faddwf1}
\mid \psi_{BC}\rangle &=& \lim_{\varepsilon\rightarrow 0}i\varepsilon
{\bf R}_{B}\mid \phi_{C}\rangle -\lim_{\varepsilon\rightarrow 0}
i\varepsilon{\bf R}_0\mid \phi_{C}\rangle \nonumber\\ &+&
\sum_{A\in {\cal C}}
\bar{\delta}_{A,B}{\bf R}_0{\bf T}_{B}\lim_{\varepsilon\rightarrow 0}
i\varepsilon{\bf R}^{A}\mid \phi_{C}\rangle\nonumber\\&=&\mid \phi_C
\rangle\delta_{B,C} - \mid\phi_{C}\rangle\delta_{C_N,C}
+\sum_{A\in {\cal C}}\bar{\delta}_{A,B}{\bf R}_0{\bf T}_{B}
\mid \psi_{AC}\rangle \nonumber\\ &=& \mid\phi_{C}\rangle\delta_{B,C}
- \mid\phi_{C}\rangle\delta_{C_N,C} +\sum_{A\in {\cal C}}
\bar{\delta}_{A,B}{\bf R}_{B}{\bf V}_{B}\mid \psi_{AC}\rangle.
\end{eqnarray}
Here the identities
\begin{equation}
\lim_{\varepsilon\rightarrow 0} i\varepsilon{\bf R}_B\mid \phi_C
\rangle =\mid\phi_{C}\rangle\delta_{B,C}
\end{equation}
and
\begin{equation}
{\bf R}_0{\bf T}_B={\bf R}_B{\bf V}_B
\end{equation}
have been used as well as the definition in Eq. (\ref{Wavef3}).  Thus
the scattering wave function, Eq. (\ref{Wavef2}), is expressed in
terms of the generalized Faddeev components (\ref{Faddwf1}),
\begin{equation}\label{Wavef4}
\mid\Psi_C\rangle=\mid\phi_C\rangle\delta_{C,C_N} +
\sum_{B\in {\cal C}}\mid \psi_{BC}\rangle.
\end{equation}
For $N$=3 this expression for the scattering state is exactly the wave
function originally derived by Faddeev.  Hence Eq. (\ref{Wavef4}) is
the generalization of Faddeev's scattering state based on a chosen set
$\cal C$ of two-cluster partitions of $N$ particles.

        It is easily shown that Eq. (\ref{Wavef4}) satisfies the
Schr\"odinger equation (\ref{Schro}).  A special case is the
homogeneous analog.  The proof that this formally satisfies the
Schr\"odinger equation is as follows:
\begin{eqnarray}\label{Wavef5}
\mid\Psi_{C} \rangle &=& \sum_{B\in {\cal C}}\mid \psi_{BC}\rangle\\
&=&\sum_{B\in {\cal C}}\left(\sum_{A\in {\cal C}}\bar{\delta}_{A,B}
{\bf R}_{B}{\bf V}_{B}\mid \psi_{AC}\rangle \right) \nonumber\\ &=&
\sum_{B\in {\cal C}}\left(\sum_{A\in {\cal C} }\bar{\delta}_{A,B}
{\bf R}_0{\bf V}_{B}\left[{\bf1}+{\bf R}_B{\bf V}_B\right]\mid\psi_{AC}
\rangle\right)\nonumber\\
&=& \sum_{B\in {\cal C}} {\bf R}_0{\bf V}_B\left(\sum_{A \in {\cal C}}
\bar{\delta}_{A,B}
\mid \psi_{AC}\rangle +\mid \psi_{BC} \rangle \right)  \nonumber\\
&=&\sum_{B\in {\cal C}}{\bf R}_0{\bf V}_{B} \sum_{A\in {\cal C}}\mid
\psi_{AC}\rangle \nonumber\\ &=&
\sum_{B\in{\cal C}}{\bf R}_0{\bf V}_B\mid \Psi_{C}\rangle \nonumber\\
&=&{\bf R}_0{\bf V}\mid \Psi_C\rangle.
\end{eqnarray}
Of course a homogeneous solution of the Schr\"odinger equation occurs
only for bound states energies, so such a solution plays no role when
solving a scattering problem.

        In general, a cluster can be either a bound, or an unbound but
interacting, set of particles, with a free particle included as a
(1-particle) bound state.  If both clusters in a 
partition  are bound  then the partition is a true
asymptotic channel.  Whether this is or is not the case, the
partition  wave function has one of two forms, depending on
whether one of the clusters is or is not a free particle, namely
\begin{equation}\label{twocluster}
\mid \phi_{C}\rangle = \cases{\mid\varphi_{C^1},{\bf q}_{C}\rangle\cr
\mid\varphi_{C^1},\varphi_{C^2},{\bf q}_{C}\rangle.\cr}
\end{equation}
Here $\mid \varphi_{C^1} \rangle$ is a bound state of the
$N-1$-particle subsystem and $\mid {\bf q}_{C}\rangle$ is the momentum
generalized eigenstate for the relative motion of the bound and free
states, on the basis that the total center of mass momentum of the
$N$-particle system has been removed from discussion.  In the second
form, the $2$nd cluster is now also a bound state, which must be
explicitly indicated, while ${\bf q}_C$ is again the relative
momentum.  If a cluster is unbound, but all particles interacting,
then the corresponding cluster wavefunction must be replaced  by
$\varphi_{C^j}^+$, corresponding to the scattering wavefunction from
some initial incoming state.  It appears intuitively reasonable that
this procedure is correct, though its complete mathematical
justification  may be needed since some  limits must
already have been taken in order to define (in general, a product of)
scattering states with definite energy as an input,  with the
next step involving a further limit.

\section{The 4- and 8-particle systems}

        The treatment so far has emphasized how the $N$-particle
scattering problem can be broken down into a number of problems
involving fewer particles.  Once a breakdown has been selected, it is
then a case of building up the transition operator and scattering
wavefunction for the $N$-particle system.  The procedure is
illustrated by discussing  systems of 4 and 8 particles. For
definiteness it is assumed that there are no bound states for any
number of particles.

        For the specific case of there being four particles, these
will be labelled 1,~2,~3 and 4.  Then according to Eq.
(\ref{noofpart}) there are 3 two-cluster partitions, which are the
three pairs of pair particles 12,34, 13,24 and 14,23.  Moreover, Eq.
(\ref{scale}) states that no scaling of the potential is required.
Thus the procedure is to first find the two-particle transition
operator ${\bf T}_{12}(z)$ for arbitrary complex $z$ in the upper half
plane.  On the basis that all particles are the same species, this
transition operator is the same for any other pair, except for the
labelling.  The second step is then to find the partition transition
operator for the two-cluster partition 12,34.  This is accomplished
according to Sec. IIIB, specifically involving either the first
approach, Eq. (\ref{Tchannel3}), or the second approach, Eq.
(\ref{TCT1T2}), with its integrand expressed in terms of the pair
particle transition operators, Eq. (\ref{integrand}).  Again this
calculation needs to be done only once since a relabelling immediately
gives the partition transition operator for the other two partitions.
The last step for getting the four-particle transition operator is to
solve the $3\times3$ matrix equations (\ref{fadd2}) and add the
results, Eq. (\ref{T_2}).  Once the transition operator is known, the
various generalized Faddeev components can be calculated, as described
in Sec. IV, and the desired scattering amplitude calculated. In this
way the four-particle scattering problem is very similar to the
three-particle scattering problem, involving only three partitions.
The extra complexity is in the added structure of the partition
transition operators.  In contrast, the method of Sloan \cite{Sloan}
involves a $7\times7$ matrix while the Yakubovsky \cite{Yakubovsky}
approach organizes the wavefunction into 18 components.  The approach
presented here would seem to be both simpler and more efficient.

        The 8-particle problem is discussed with the view of
clarifying how the proposed approach works for more complicated
systems. Equation (\ref{noofpart}) states that there are 35 two-cluster
partitions, each cluster having 4 particles.  Moreover, Eq.
(\ref{scale}) states that $v_8=1/15$, so the potential needs to be
scaled by this fraction.  This scaling is required when finally
solving the $35\times35$ matrix, Eq. (\ref{fadd2}), for the components
of the 8-particle transition operator, so enters into the
calculation of all (the 8-particle, partition, cluster and
subpartition and subcluster) transition operators.  Now a typical
8-particle partition transition operator is needed, which is
a combination of two 4-particle cluster transition operators.  The
latter must be found by solving the problem discussed in the previous
paragraph, but now with the potential scaled by $v_8$.  Since there is
no further scaling required in solving for the 4-particle transition
operator, it is only the $v_8$ scaling that must be applied. 
Thus the detailed  procedure that is to be followed is, 
in sequence: 1) scale the pair potential by $v_8$; 2) calculate
the 4-particle transition operator for this scaled potential as
described in the last paragraph; 3) use either of Eqs.
(\ref{Tchannel3}) or (\ref{TCT1T2}) to obtain a two-cluster partition
transition operator for the 8-particle system, with the second cluster
transition operator obtained from the first 4-particle cluster
transition operator by relabelling; 4) obtain the 34 other partition
transition operators by relabelling and solve the generalized Faddeev
equations (\ref{fadd2}).

\section{DISCUSSION}

        A proposal has been made for the generalization of the Faddeev
equations\cite{Faddeev} to an arbitrary number, $N$, of particles.
Essentially this involves selecting a set of partitionings of the $N$
particles into pairs of clusters and expressing the scattering
properties of the total system in terms of the scattering properties
of the clusters.  Each cluster is treated in the same way so that
after successive treatments an individual cluster contains either only
a single particle or a pair of particles.  Although mathematically,
other (combinations of) sets of two-cluster partitions lead to a
connected kernel as well in the present method, the detailed
presentation of this paper has stressed the choice of the set of
two-cluster partitions in which the pair of clusters in any partition
is of nearly equal size.  This should be the most efficient procedure
since then there are fewer types of clusters that need to be treated.

        The proposed method requires, in general, a scaling of the
potential and expressing the transition operator, and wave function, in
terms of partition transition operators which are in turn expressed in
terms of the transition operators for the two individual clusters
composing the partition.  An essential simplifying feature is that
the transition operator for an individual cluster can be calculated
independently of the presence of the other particles, so is equivalent
to solving the scattering of a system of fewer particles, which can
in general be treated in the same manner as the original system.

       For most $N$-particle integral equation theories, the starting
point is the pair particle interaction or the two-particle transition
operator, which is then combined in all possible ways with the
transition operators of other pairs of particles.  In order to assure
that all particles are connected, that is, no disconnected diagrams
appear, it is necessary to iterate through a whole sequence of
processes.  This complicates the description of the $N$-particle
problem as it appears, for example, in the Yakubovsky
\cite{Yakubovsky} equations.  In contrast, the present method starts
by dividing the $N$-particle system into a selected set of two-cluster
partitions.  The treatment assures that all clusters are connected to
each other.  There remains the possibility that there is a
disconnectionness within a cluster.  This is eliminated by repeating
the procedure for each cluster as if it were a separate scattering
system. In this way the formulation has a very simple structure which
is exactly similar at each stage to the Faddeev equations, in
particular reducing to them if $N=3$.

       The total resolvent for the $N$-particle system is decomposed
in a manner analogous to Faddeev's method.  This decomposition
determines the decomposition of the scattering wave functions, viz.
Eq. (\ref{Wavef4}), into what are here called generalized Faddeev
components.  The sum of the corresponding homogeneous set of
generalized Faddeev components is explicitly shown to be completely
equivalent to the Schr\"{o}dinger equation.  Such a property was
explicitly demonstrated in Faddeev's theory\cite{Glockle2} and
Yakubovsky's theory\cite{Yakubovsky} but has not been explicitly shown
in some other theories.  This property is of particular importance for
finding the (discrete) eigenvalues for the $N$-particle system and
for avoiding spurious solutions (for bosons and fermions, an
appropriate symmetrization of the Faddeev components may be required).
In particular, Federbush\cite{Federbush} first found a spurious
solution for the Weinberg equation\cite{Weinb} and a systematic study
by Gl\"{o}ckle and coworkers\cite{Glockle3,Glockle4} indicated that
most few-body equations admit the existence of discrete spurious
solutions.  However these spurious solutions were not a problem in
finding scattering solutions.

        The cluster and partition wavefunctions from all the stages
are needed before the total wavefunction can be obtained.  In general
these are nonphysical in that they represent intermediary results for
the final calculation.  The same is true even more so for the Faddeev
components.  In contrast, if both clusters in a partition represent
bound states, then this is a valid asymptotic condition and the
channel wavefunction does represent a physical state for the
corresponding scattering system.

\section*{ACKNOWLEDGMENTS}

The authors thank the referee for drawing their attention to the
method of Haberzettl and Sandhas for combining dynamically independent
transition operators, which procedure has been 
incorporated into the presentation in Sec. IIIB. This work
was supported in part by the Natural Sciences and Engineering Research
Council of Canada. G.W.W. thanks the Killam Foundation of Canada for a
fellowship and the National University of Singapore for research
funds.

\end{document}